\documentstyle[preprint,revtex]{aps}

\begin{document}
\draft

\begin{title}
Path Integral Monte Carlo Study \\
of a Model 2D Quantum Paraelectric.
\end{title}

\author{R. Marto\v{n}\'{a}k, E. Tosatti\cite{et}}

\begin{instit}
International School for Advanced Studies (S.I.S.S.A.) \\
Via Beirut 2--4, 34014 Trieste, Italy
\end{instit}

\begin{abstract}
We have begun a study of quantum ferroelectrics and paraelectrics. Simple 2D
short-range lattice model hamiltonians are constructed, keeping in mind the
phenomenology of real perovskite systems, like $SrTiO_{3}$ and $KTaO_{3}$.
Pertinent quantum tunneling phenomena, and the presence of an ice-like
constraint are demonstrated. The two simplest models, namely a plain quantum
four-state clock model, and a constrained one, are then studied in some detail.
We show the equivalence of the former, but not of the latter, to a quantum
Ising
model. For the latter, we describe a very good analytical wavefunction valid in
the special case of zero coupling ($J = 0$). In order to study the full quantum
statistical mechanics of both models, a Path Integral Monte Carlo calculation
is
set up, and implemented with a technique, which even in the constrained case
permits a good convergence for increasing time slice number $m$.
The method is applied first to the unconstrained model, which serves as a
check, and successively to the constrained quantum four-state clock model.
It is found that in both cases, a quantum phase transition still takes place at
finite coupling J, between a ferroelectric and a quantum paraelectric state,
even when the constraint hinders disordering of the ferroelectric state.
This model paraelectric state has a finite excitation gap,
and no broken symmetry. The possible role of additional ("oxygen hopping")
kinetic terms in making closer contact with the known phenomenology of
$SrTiO_{3}$ is discussed.

\end{abstract}


\section{Introduction.}

It is well-known that large enough quantum fluctuations can de-stabilize the
ferroelectric state in favour of a paraelectric state, which possesses no net
polarization order parameter even at zero temperature. \cite{mul1}.
Unlike classical paraelectrics, where
disorder is temperature-induced, quantum paraelectrics (QPE) are in a sense
very
much ordered, in that their ground state corresponds to a well-defined
wavefunction
encompassing all dipoles in the macroscopic system. This aspect makes them
especially interesting, also in connection with a recent suggestion \cite{mbt},
that some kind of macroscopic coherence, or ODLRO, might appear under special
circumstances.

Three-dimensional examples of quantum paraelectric states are believed to be
provided, among others, by the low-temperature states of $SrTiO_3$
\cite{mul1,mbt,mul2}, $KTaO_3$ \cite{rytz}, as well as $KH_{2}PO_4$ above
16.6 Kbars \cite{sam}. In these
systems, quantum fluctuations suppress ferroelectricity, characterized by an
$n$-component real order parameter ($n=2$ in tetragonal $SrTiO_3$, $n=3$ in
cubic $KTaO_3$, $n=1$ in $KH_{2}PO_4$).

Interest in quantum ferroelectrics started in the late 50's \cite{blinc}, and
revived
again in the late 70's \cite{mul1}, at a time when attention was focused on
critical exponents of the quantum phase transition, including the crossover
between quantum and classical critical exponents \cite{opp,schneid,morf}.
It has continued more recently, with impurity-induced transitions from
quantum paraelectric to domain-type ferroelectric \cite{bm,rod,kleem}. Apart
from that, there seems to exist no real attempt as yet to study the nature of
the quantum-mechanical state of QPE's in a well defined realization.

Our goal with this paper is to begin a microscopic study of models which
exhibit
quantum ferroelectric and quantum paraelectric behaviour. The method we will
apply is a mixture of variational, mean-field theory, and principally
Path Integral Monte Carlo (PIMC), coupled when possible with finite
size scaling \cite{barber} .

The first, and main problem, is the choice of an appropriate model hamiltonian.
In choosing the model, we try to adhere as much as possible to the actual
physical situation of $SrTiO_3$, whose phenomenology seems best known. However,
a drastic amount of simplification is clearly needed. On one hand, the real
crystal is rich of complications, such as coupling to antiferrodistortive and
to elastic degrees of freedom \cite{uwesakudo}. On the other hand, even if PIMC
is a powerful tool for quantum statistical mechanics, the system one can hope
to
study with present means is by necessity very simple and relatively
small-sized.
Our first step, therefore, must be a judicious choice of the main ingredients
in the problem. We can be guided, first of all, by existing understanding of
classical ferroelectrics.

The plan of this paper is as follows. In section 2 we first briefly review the
phenomenology of classical ferroelectrics, and quantum paraelectrics, with
emphasis on the crossover from the displacive to the order-disorder regime.
After demonstrating the relevance of the four-state clock model in the case of
$SrTiO_3$, we discuss in detail various possible quantum effects, as well as
the presence of several important constraints. In section 3 we then consider
the simple (unconstrained) quantum four-state clock model. Its general
equivalence to two uncoupled Ising models, already known classically, will be
proven for the quantum case as well. In Section 4, we add a constraint to the
quantum four-state clock model, and discuss a simple approximate wavefunction
for the case of zero temperature and zero coupling. Section 5 introduces the
technicalities of our PIMC
calculation. In order to treat the constraint properly and preserve the usual
$1/m^2$ convergence in the number of Trotter slices, we introduce a particular
version of the checkerboard decomposition scheme, which is commonly used, e.g.
for simulation of quantum spin systems \cite{tan}. Section 6 presents a PIMC
test study of the simple quantum
four-state clock model. Comparison with the known results of the quantum
Ising model, done throughout that section, serves as a check of the basic
soundness of the PIMC method, as well as of the kind of errors to be expected
in
later applications. In section 7 the PIMC is applied to the {\it constrained}
four-state clock model. It is found, as expected, that for given values of
quantum hopping and coupling constant,
ferroelectricity is now stronger, and the ferro-para transition occurs at much
higher temperature in the constrained case than in the unconstrained case.
In spite of this strengthening of ferroelectricity, there still is a critical
value of the coupling constant, below which the system is a quantum
paraelectric
at $T=0$. However, this critical value is now a
factor of $\sim 4$ smaller, due to the constraint. Finally, Section 8 is
devoted to a general discussion, and relevance to the known phenomenology of
the
perovskite QPE's. In particular, the $T=0$ state of both models studied so far
does not exhibit in the QPE regime any broken symmetry or ODLRO, and can be
reached from the high temperature paraelectric state without a phase
transition.
Therefore, the question of understanding the phase transition described by
\cite{mbt} is left open.

\section{Choice of models.}

Microscopic models for classical ferroelectricity have been available
for a long time \cite{bruce}. Among them, very popular are the soft-mode
lattice-dynamical models \cite{nato}, and the ice models \cite{slater}.
Ferroelectricity, in fact, represents just a particular case of a more general
class of phenomena, structural phase transitions.
An extensive and definitive review of critical phenomena related to classical
structural transitions has been provided in Ref. \cite{bruce}. Here are a few
pertinent points.

Starting with a generic model, which consists of a lattice of particles in a
local double-well potential, and connected by harmonic interparticle springs,
a structural transition problem will approach, depending on parameters, one of
two opposite regimes: the {\it displacive} or the {\it order-disorder} regime.
In the displacive limit, fluctuations are unimportant, whence the local
distortion and the global order parameter can be identified. They are finite
below the critical temperature  $T_{c}$ and zero above $T_{c}$. The appropriate
description in this case is the standard soft-mode mean-field theory
\cite{bruce}, based on continuous degrees of freedom.

In the opposite, order-disorder limit, fluctuations dominate. The local
distortion order parameter is generally large, fluctuating, and very different,
as temperature grows, from the global order parameter.
In particular, the former evolves smoothly with
temperature, while the latter of course vanishes critically at $T_{c}$. If one
further assumes that close to $T_{c}$ the local distortions remain exactly
constant as $T$ varies, then the continuous degrees of freedom can be replaced
without harm by a discrete variable.

For a general classical structural transition, it is well-known that close
enough to $T_{c}$ the critical behaviour for dimension $d < 4$ is
fluctuation-dominated, at least with short-range forces. In other words, even a
system whose behaviour is displacive well below and well above $T_{c}$, will
undergo a crossover into an order-disorder regime, close enough to, and on both
sides of, $T_{c}$. The critical temperature region is characterized by the
appearance of fluctuating ordered domains of large size $\xi$. Inside the
domains the order parameter is large, and locally coherent. However, different
domains are incoherent \cite{schst}.
The order parameter dynamics shows in this regime none of the softening
typical of the displacive limit. Critical dynamics is instead due to the slow,
sluggish relaxation of the domain walls, leading to the famous "central peak"
well known from Raman scattering \cite{bruce}.

In a ferroelectric, fluctuating domains give rise to a typical, very slow,
Debye-like contribution to the low-frequency dielectric susceptibility, of the
form $\Delta \epsilon(T)/(1 + i \omega \tau(T))$. If the ferroelectric
transition is continuous, both $\Delta \epsilon$ and $\tau$ diverge at $T_{c}$.
The critical exponents of the real system are in the same universality class of
a completely discrete model, provided the two have the same symmetry. A
discrete
model is therefore sufficient to capture the basic physics of the classical
ferroelectric-paraelectric transition.

What about a quantum paraelectric ? At high enough temperature, the quantum
perovskites $SrTiO_{3}$, $KTaO_{3}$, become just ordinary classical
paraelectrics, well described by the displacive limit. This classical
displacive
behaviour is confirmed by IR and Raman spectra showing very well-defined TO
modes, hard and narrow. These modes soften upon cooling, as expected in the
displacive picture \cite{lyons}. However, just above the extrapolated classical
Curie temperature $T^{*}$ (37 K for $SrTiO_{3}$, 40 K for $KTaO_{3}$), the IR
active modes broaden dramatically. At the same time, the slow Debye relaxations
typical of the order-disorder regime appear in the microwave region, their
typical frequency $\tau^{-1}$ again decreasing with temperature
\cite{maglione}.
This is the usual signal of crossover into the critical, order-disorder regime
of ordinary classical paraelectrics. {\it This crossover therefore takes place
also in quantum paraelectrics.} Unlike the classical systems, however, in this
case the critical slowing down, i.e. the divergence of $\tau$ with decreasing
$T$, appears to be blocked at some {\it finite}
relaxation time $\tau^{*}$, which in both $SrTiO_{3}$ and $KTaO_{3}$ is very
long, $(\tau^{*})^{-1} < \sim 500 MHz$ \cite{maglione1}. Because of this lack
of
divergence of $\tau$, long-range ferroelectric order is never reached, and the
system remains paraelectric even at the lowest temperatures. The failure to
order ferroelectrically accompanied by these slow dielectric fluctuations has
been attributed to quantum zero-point motion \cite{mul1}. There is clear
evidence
showing that fluctuations can be easily removed, the system correspondingly
turning into a regular ordered ferroelectric, by applying either pressure
\cite{uwesakudo}, electric field \cite{fleury}, and impurity doping \cite{bm}.
We conclude that when approaching $T^{*}$ from above, the QPE perovskites are
well inside the order-disorder regime. A sort of quantum order-disorder regime
appears moreover to persist all the way down to $T=0$. Between $T^{*}$ and
$T=0$, we have a kind of "quantum central peak" state: - rather than completing
its classical slowing down, and undergoing a regular ferroelectric critical
point transition, the system hangs indefinitely on the verge of criticality,
due to quantum fluctuations.

An important corollary of this discussion is that a discrete lattice model,
which lends itself better to a description of order-disorder critical
fluctuations, is more likely to yield, probably even in the details, a better
description of the QPE state, than a continuous, displacive model.

The paradigmatic discrete model
for quantum paraelectricity with a scalar order parameter is the Ising model
in transverse field \cite{degen}, \cite{elliott}, which was extensively studied
in the 70's, using a variety of techniques \cite{pfeuty}. Its relevance is
particularly direct for the description of hydrogen-bonded ferroelectrics, like
$KH_{2}PO_4$. The order parameters in $KTaO_3$ and $SrTiO_3$ are not exactly
Ising-like, however.
$KTaO_3$ remains cubic down to the lowest temperatures and the order parameter
has thus three components. $SrTiO_3$, which acquires a tetragonal structure at
low temperatures (below 105 K), can become ferroelectric when doped by $Ca$
and the resulting ferroelectricity is known to be XY-like \cite{bm}. In pure
$SrTiO_3$, in particular, if the tetragonal axis $z$ is taken along $(001)$,
so that below 105 K two neighbouring $TiO_3$ octahedra rotate by an angle
$\Phi$ and $-\Phi$ around $z$ ($\langle \Phi \rangle \simeq 5^{\circ} $ well
below 105 K), then ferroelectricity shows a tendency to occur only along either
$\pm(100)$ or $\pm(010)$, and more generally in the (001) $(x,y)$ plane.
In the ferroelectric state, the $Ti$ central ion and one of the four coplanar
oxygens which surround it in the $(x,y)$ plane, establish among themselves a
{\it slightly stronger} bond than the other three. Coupling between different
cells is not exclusively dipole-dipole, but should have important electronic
contributions (short-range), and elastic contributions (long-range). The
elastic
coupling
mechanism, in particular, is likely to be very strong, as confirmed by the fact
that even a small pressure along $x$ is sufficient to yield ferroelectric order
along $y$ \cite{uwesakudo}.
Physically, what we believe happens, is that a double well situation for the
central $Ti$ ion, or equivalently for a bridging oxygen between two $Ti$ ions,
say along $(100)$, can occur only if the lattice is {\it locally} expanded
along that direction. The consequences of coupling of ferroelectricity with
elastic modes, leading in particular to incommensurability phenomena, and the
possible role of quantum effects have been given a separate discussion
elsewhere
\cite{rm}.

In the present lattice modelling, we shall however ignore these details, and
assume simply a short-range ferroelectric coupling $J$, between first neighbour
cells. Each cell has an $XY$ phase variable $\phi$, representing the $Ti$
displacement, or the dipole direction, subject to a cubic anisotropy, with
minima at $\phi = 0, \pm \pi/2, \pi$. In the large anisotropy limit, a minimal
lattice model of ferroelectricity in $SrTiO_3$ is therefore a 3D four-state
clock model with first neighbour coupling. This leaves entirely out possible
long-range effects due to dipole-dipole coupling, and to coupling to elastic
modes, as mentioned above, as well as additional coupling to the
antiferrodistortive order parameter. In order to make quantitative progress, we
choose to ignore all these complications for the time being,
even though they will probably have to be reconsidered at a later stage.

By introducing at each site the complex variable
\begin{equation}
z_{i} = e^{i \phi_{i}} \; , \label{complex}
\end{equation}
where $\phi_{i}$ can take the four values $0, \pm \pi/2, \pi$, the ordinary,
classical four-state clock model can be written as
\begin{eqnarray}
H^{4} &=& \sum_{i} H^{4}_{i} \, \nonumber\\
H^{4}_{i} &=& - {{J}\over{2}} \sum_{j}\cos (\phi_{i} - \phi_{j})
= - {{J}\over{2}} \mbox{Re}(\sum_{j} z_{i} z^{*}_{j}) \, , \label{cl4}
\end{eqnarray}
where the sum over $j$ runs over nearest neighbours of $i$.

It was shown by Suzuki \cite{suzuki} that the classical four-state clock model
is mapped in full generality, i.e. independently both of dimensionality, and of
range of interaction, onto two decoupled Ising models. Hence, everything is
known about the classical behaviour of model (\ref{cl4}).

However, we must now introduce quantum mechanical effects, in the form of a
kinetic energy term, not commuting with the potential energy (\ref{cl4}). In a
perovskite ferroelectric, one can envisage at least two distinct quantum
effects, both resulting in dipole tunneling.

The first is quantum hopping of the central positive ion (which will be called
$Ti$, since it is $Ti$ in $SrTiO_3$), from bonding preferentially one oxygen to
bonding the next one, within the same cage, or cell [Fig.1]. The kinetic energy
for this is, in the continuous case
\begin{equation}
H^{kin\,1} = {{\mu\rho^{2}}\over{2}} \sum_{j} \dot{\phi}_{j}^{2} =
{{1}\over{2\mu\rho^{2}}} \sum_{j} \left(-i \hbar {{\partial}\over
{\partial\phi_{j}}} \right)^{2} \; , \label{kin1}
\end{equation}
where $\mu$ and $\rho$ are effective mass and off-center displacement (a very
small quantity, of order of $0.03 \AA$ \cite{mul1}) of the $Ti$ ion.
In our discrete case, we shall allow the clock variable $z_{j}$ to hop, for
simplicity, into its two nearest orientations, i.e. from $z_{j}$ into
$\pm i z_{j}$. If we choose to describe the system with a wavefunction
$\Psi(z_{1},\ldots,z_{n})$, the corresponding kinetic piece of hamiltonian
reads
\begin{eqnarray}
H^{kin\,1} &=& \sum_{j} H^{kin\,1}_{j} \; \nonumber\\
H^{kin\,1}_{j} \Psi(z_{1},\ldots,z_{j},\ldots,z_{n})
&=& -t ( \Psi(z_{1},\ldots,i z_{j},\ldots,z_{n}) +
\Psi(z_{1},\ldots,-i z_{j},\ldots,z_{n}) ) \; . \label{disc}
\end{eqnarray}
In the last expression, the hopping energy has further been lumped into the
constant $t = {{\hbar^{2}}\over{2\mu\rho^{2}}}$. This term is an obvious
generalization of the "transverse field" term in the Ising case. If strong
enough, it will
cause the dipole in each cell to hop quantum mechanically from one value of
$\phi_{i}$ to the other, irrespective of the state of dipoles in neighbouring
cells, thereby destroying ferroelectric long-range order.

A second type of quantum effects, which has apparently not been discussed so
far, is {\it oxygen double-well tunneling}. When an oxygen is bonded to a given
$Ti$ ion, we can imagine another energy minimum, more or less equivalent, when
that oxygen is bonded to the other $Ti$ ion on the opposite side [Fig.2.].
Quantum hopping processes of the oxygen between these two sites (similar to
proton hopping in hydrogen bonds) may play a role in quantum paraelectrics,
both because of a relatively small displacement involved, and of the small
oxygen mass. However, these processes cause the bond to hop from one cage to
the next and therefore, unlike the intracage hopping $t$, they do not obey
the apparently sensible requirement that the number of bonds per cage should
not
exceed one. An adequate prescription is thus necessary to include the oxygen
tunneling into our scheme.

We have considered so far two possible implementations of oxygen tunneling.
The first is one where we force each cage to retain {\it one and only one
dipole
bond} in any given configuration. This is probably close to what happens at
very
low temperatures. The second is a different scheme, where we introduce the
possibility of "bond vacancies", while still forbidding double bond occupancy
of any cage. We discuss both schemes in this order.

If each cage is constrained to retain strictly one and only one bond at any
given time, bond hopping becomes extremely problematic. For each bond which
hops, a suitable "bond backflow" loop is mandatory, so as to obey the
constraint
everywhere. In other words, only {\it concerted rearrangements}, such as
ring-shaped currents of bonds, are permitted. The simplest concerted bond
hopping loop is shown in Fig.3. The associated kinetic energy is a one-body
operator, which for a horizontal link $\langle i,j=i+\vec{x} \rangle$ can be
symbolically written as
\begin{equation}
H^{kin\,2a}_{ij} = -t^{'} \left(
| \leftarrow_{j} \rangle \langle \rightarrow_{i} | +
| \rightarrow_{i} \rangle \langle \leftarrow_{j} | \right) \; , \label{kin2a}
\end{equation}
and analogously for a vertical link. In process (\ref{kin2a}), concerted ring
bond hopping takes place on the elementary plaquette only. Also, it is
ineffective if the initial state is a fully aligned ferroelectric. More general
concerted processes should also be interesting, and do not necessarily have
this
limitation. However, we do not wish to go into this discussion at this stage.

The second possibility we envisage, having particularly in mind the situation
at
not too low temperatures, is one where we still forbid multiple bond occupancy
of a cage, but we allow for the possibility of {\it bond vacancies}, i.e. cages
where the $Ti$ ion is centrally located, and has no dipole bond at all. The
justification for this comes from the necessity to include at temperatures
close
to $T^{*}$ and higher, elements which link the present discrete model with the
displacive picture, eventually applicable when $T \gg T^{*}$. At these very
high
temperatures, the likeliest $Ti$ ion location is indeed the cage center, and
there are physically no bonds to speak of. Introduction of some bond vacancies
even at lower temperatures allows for a new bond hopping process, where a
vacancy on one cage and a bond on a neighbouring cage can exchange positions.
The link oxygen between two cages can "resonate" between two adjacent $Ti$
ions,
both of which are free from other bonds. The corresponding piece of hamiltonian
reads symbolically, again for a horizontal link
\begin{equation}
H^{kin\,2b}_{ij} = -t^{'} \left( | \rightarrow_{i} \circ_{j} \rangle
\langle \circ_{i} \leftarrow_{j} | + {\mbox h.c.} \right) \; , \label{kin2b}
\end{equation}
and analogously for a vertical link, and $\circ_{i}$ denotes a vacancy on site
$i$. Through processes (\ref{kin2b}), vacancies and bonds become mobile, and
the
possibility of condensates may arise at sufficiently low temperatures.

In this initial paper, we will restrict to consider only the on-site kinetic
energy (\ref{disc}), as a first step. Dealing with processes of the type
(\ref{kin2a}) or (\ref{kin2b}) is evidently more complex, and we defer that
until after obtaining a good understanding of the more basic terms.

Besides potential and kinetic energies, we have found that the problem of
ferroelectricity (both classical and quantum) in a perovskite requires a
third, novel ingredient. That ingredient is an anholonomic {\it constraint},
which in hamiltonian terms can be mimicked by some {\it infinite} repulsion, or
attraction. The physical constraint we consider is that no oxygen should be
bound (i.e. form a dipole) simultaneously to both $Ti$ atoms on the two sides
of
the link where it belongs. In other words, while a $Ti$ atom has always one and
only one bond (barring vacancies from now on), an $O$ atom can have either zero
or one bond, but not two: the configuration of Fig.4 is physically meaningless,
and should not occur. The constraint can also be included as a hamiltonian
term,
in the following way. First we introduce for each link an operator
\begin{equation}
P_{ij}^{'} = \delta( z_{i} - r_{ij} ) \delta( z_{j} + r_{ij} ) \; ,
\end{equation}
which acts as a projector on the forbidden states. In this expression,
$r_{ij} = r_{j} - r_{i}$, where $r_{i}$ is the complex number defining the 2D
position of site $i$. The hamiltonian term equivalent to the constraint then
reads
\begin{equation}
H^{constr} = \lim_{U \rightarrow \infty }
U \sum_{\langle ij \rangle} P_{ij}^{'} \; . \label{constr}
\end{equation}
Enforcement of this constraint is analogous to an "ice rule", and has also
apparently never been considered before for the perovskites. It is likely to
yield nontrivial modifications in the physics. At the classical level, for
example, we expect a dramatic reduction of entropy due to the decreased number
of available configurations, and a possible change of the critical point
universality class. Related effects are expected in the quantum paraelectric
problem, which is of direct interest to us.

Finally, we have to choose dimensionality. While the physical system is of
course in $D=3$, we see no basic harm in restricting our study to a $D=2$
square
lattice of cages. This amounts to treating, so to speak, a single $TiO_{2}$
plane, ignoring interplanar coupling. This seems a particularly plausible
approximation in the case of $SrTiO_{3}$, where ferroelectric $Ti$ -- $O$ bonds
are strictly planar, and do not involve the interplanar bridging oxygens.

\section {\bf Simple Quantum Four-State Clock model.}

In this section we shall consider the simplest of the models, which is
defined by the hamiltonian
\begin{equation}
H = H^{4} + H^{kin\,1} = \sum_{i}( H^{4}_{i} + H^{kin\,1}_{i} ) \; .
\label{simple}
\end{equation}
This model can be mapped on two decoupled Ising models in a
transverse field, regardless of the dimensionality.
In order to construct the
mapping, it is convenient to start from the $\phi_{i}$ representation.

We represent the clock variable $\phi_{i}$ on each site by two discrete
variables, $s_{i}$ and $\sigma_{i}$, defined respectively by
\begin{eqnarray}
s_{i} &=& \sqrt{2} \cos(\phi_{i} + {{\pi}\over{4}}) \nonumber\\
\sigma_{i} &=& \sqrt{2} \sin(\phi_{i} + {{\pi}\over{4}}) \; . \label{ssigma}
\end{eqnarray}
It is easily seen that $s_{i}, \sigma_{i} = \pm 1$, and thus the new variables
can be regarded as Ising variables. The potential energy term can be
immediately
written as
\begin{equation}
H^{4}_{i} = - {{J}\over{2}} \sum_{j}\cos (\phi_{i} - \phi_{j}) =
- {{ {{1}\over{2}} J}\over{2}} \sum_{j} (s_{i} s_{j} + \sigma_{i} \sigma_{j})
\; ,
\end{equation}
which is just Suzuki's decoupling for the classical case \cite{suzuki}.

The on-site hopping term can be written as
\begin{eqnarray}
& &H^{kin\,1}_{i} \Psi(s_{1},\sigma_{1},\ldots,s_{i},\sigma_{i},\ldots,
s_{n},\sigma_{n}) =
H^{kin\,1}_{i} \Psi(\phi_{1},\ldots,\phi_{i},\ldots,\phi_{n}) =\nonumber\\
& & -t \left( \Psi(\phi_{1},\ldots,\phi_{i} - {{\pi}\over{2}},\ldots,\phi_{n})
+
\Psi(\phi_{1},\ldots,\phi_{i} + {{\pi}\over{2}},\ldots,\phi_{n}) \right) =
\nonumber\\
& & -t \left( \Psi(s_{1},\sigma_{1},\ldots,\sigma_{i},-s_{i},\ldots,s_{n},
\sigma_{n}) + \Psi(s_{1},\sigma_{1},\ldots,-\sigma_{i},s_{i},\ldots,s_{n},
\sigma_{n}) \right) \; .
\end{eqnarray}
Because $s_{i}, \sigma_{i} = \pm 1$, we have always $s_{i} = \pm \sigma_{i}$,
and the last expression can thus be rewritten as
\begin{eqnarray}
& &H^{kin\,1}_{i}
\Psi(s_{1},\sigma_{1},\ldots,s_{i},\sigma_{i},\ldots,s_{n},\sigma_{n}) =
\nonumber\\
& & -t \left( \Psi(s_{1},\sigma_{1},\ldots,-s_{i},\sigma_{i},\ldots,s_{n},
\sigma_{n}) + \Psi(s_{1},\sigma_{1},\ldots,s_{i},-\sigma_{i},\ldots,s_{n},
\sigma_{n}) \right) \; .
\end{eqnarray}
We thus see that the Ising variables $s_{i}, \sigma_{i}$ are decoupled both in
$H^{4}_{i}$ and in $H^{kin\,1}_{i}$, and the latter term corresponds to the
well-known transverse field. The quantum four-state clock model is thus
equivalent to two {\it decoupled} Ising models in a
transverse field equal to the original hopping parameter $t$, and with a new
coupling constant $J/2$.

The Ising model in a transverse field was intensively studied in the 70's. An
exact solution for 1D case can be found in Ref. \cite{pfeuty}. Discussions of
higher dimensionalities are given in several places
\cite{moore1,moore2,moore3}.
Quantum Monte Carlo renormalization group study of 1D and 2D case is available
in Ref. \cite{kolb}. In 2D
at $T=0$, there is a continuous quantum phase transition at $J_{q}/t = 0.33$
between a ferroelectric state and a QPE state. The critical point extends into
a line at finite temperature, where a characteristic pattern of
quantum-classical crossover is realized \cite{pfeuty1}.

We shall come back to the simple quantum four-state clock model once again in
this paper, in section 6, where we use this model as a test case for our Path
Integral Monte Carlo algorithm.

\section {\bf The constrained quantum four-state clock model.}

In this section we shall consider the constrained model, defined by the
hamiltonian
\begin{equation}
H^{tot} = H^{4} + H^{kin\,1} + H^{constr} \; . \label{ham}
\end{equation}

First of all we notice that the constraint eliminates very large number of
configurations. To get an insight, let us consider a 1D case for a moment, and
for the sake of simplicity let us force the constraint on each other link only
(Fig.5). Clearly, the number of allowed configurations in this simplified
case represents an upper limit of the number of allowed configurations in the
actual 2D case.

Let us denote the total number of configurations in the chain of $2 N$ sites as
$b(N)$, and that of forbidden configurations as $a(N)$. Adding two new sites to
the chain (increasing N by 1) amounts to increasing $b(N)$
by a factor of $4^{2}$. The sequence $b(N)$ is therefore defined by the
relation
$b(N + 1) = 16 \, b(N)$, plus the boundary condition $b(1) = 16$, which yields
$b(N) = 16^N$.

An analogous recursion formula can be found expressing the number of
forbidden configurations $a(N + 1)$ in the chain of $2(N + 1)$ sites through
the number $a(N)$ in the chain of $2 N$ sites. To do so, we notice that a
configuration of the chain of $2(N + 1)$ sites is forbidden when {\it either}
the two additional
sites are in their single forbidden configuration while the original $2 N$
sites are in any of the total of their $b(N)$ configurations, {\it or} when the
additional sites are in any of their 15 allowed configurations, but the
original
$2 N$ sites are in one of their $a(N)$ forbidden configurations. The sequence
$a(N)$ therefore satisfies the recursion formula $a(N + 1) = 15 a(N) + b(N)$
with boundary condition $a(1) = 1$. The explicit expression for $a(N)$ is
therefore
$a(N) = 16^{N-1} + 15 \times 16^{N-2} + \ldots + 15^{N-2} \times 16 +
15^{N-1}$.

We are interested in the ratio ${{a(N)}\over{b(N)}}$ and its limit as
$N \rightarrow \infty$. Using the above expressions for $a(N)$ and $b(N)$ we
find that ${{a(N)}\over{b(N)}} = 1 - ({{15}\over{16}})^{N} \rightarrow 1$, as
${N \rightarrow \infty}$, which means that in the thermodynamic limit {\it the
constraint eliminates "all" the configurations}, except for a set of zero
measure. More precisely, the ratio of dimensionality of the
constrained Hilbert space to that of the unconstrained one goes to zero in the
thermodynamic limit, which means that the constrained space is {\it orthogonal}
to the original one \cite{foot1}.

To start with, we shall investigate the properties of the constrained
four-state
clock model for the specially simple case $J = 0$. In this limit, it is very
easily checked that the ground state of the unconstrained system
($U = 0$) is nondegenerate, and is described by a wavefunction $\Psi_{g}^{0}$,
which is just a constant. This state can be seen as the product of $N$ separate
identical wavefunctions for each cell, each corresponding to the $j = 0$
angular
momentum state of that cell. The corresponding many-body first excited state is
infinitely degenerate. If we choose
to label the wavefunctions by a wavevector $\vec{k}$, we have for each
$\vec{k}$
two independent states, described respectively by wavefunctions (unnormalized)
\begin{eqnarray}
\Psi_{exc1}^{0}(\vec{k},z_{1},\ldots,z_{n}) &=& \sum_{i}
e^{i \vec{k} \vec{R}_{i}} \mbox{Re} \, z_{i} \Psi_{g}^{0} \label{exc0} \\
\Psi_{exc2}^{0}(\vec{k},z_{1},\ldots,z_{n}) &=& \sum_{i}
e^{i \vec{k} \vec{R}_{i}} \mbox{Im} \, z_{i} \Psi_{g}^{0} \; ,
\end{eqnarray}
where $\vec{R}_{i}$ is the ordinary (real) vector defining the position of
site $i$. In the absence of the constraint, i.e. for $U = 0$, the first excited
state is separated from the ground state by an energy gap $2 t$ (in the rest
of the paper we shall always set $t = 1$).

What about the constrained  $U = \infty$ model ? Guided by solving first the
two-site problem, we try as an ansatz for the ground state wavefunction of the
constrained system the following (Jastrow-like) product state
\begin{equation}
\Psi_{g}^{\infty}(z_{1},\ldots,z_{n}) = \prod_{\langle ij \rangle}
( \mid z_{i} - r_{ij} \mid + \mid z_{j} + r_{ij} \mid ) \; , \label{ansatz}
\end{equation}
which contains no free parameters.

We shall also assume that the first excited state of the constrained system
corresponds to $\vec{k} = 0$ and can be obtained as the analytic continuation
of
the corresponding state (\ref{exc0}) from $U = 0$ to $U = \infty$. In this way
we arrive at the wavefunction
\begin{equation}
\Psi_{exc1}^{\infty}(\vec{k}=0,z_{1},\ldots,z_{n}) = \sum_{i} \mbox{Re} \,
z_{i}
\Psi_{g}^{\infty}(z_{1},\ldots,z_{n}) \; , \label{excon}
\end{equation}
and analogously for the degenerate state $\Psi_{exc2}^{\infty}$.

To test these guessed wavefunctions in the real 2D case, defined by the
constrained hamiltonian (\ref{ham}), we have performed a diagonalization
for a $2 \times 2$ and a $3 \times 3$ system with periodic boundary conditions
(Lanczos diagonalization). In Table 1 we show the overlap of trial states
(\ref{ansatz}) and (\ref{excon}) with the exact wavefunctions, and
compare the corresponding ground state and excitation energies. It turns out
that (\ref{ansatz}) is an excellent approximation for the ground state of these
small systems, and the same is true for the excited state ansatz (\ref{excon}).
We feel therefore encouraged to adopt (\ref{ansatz}) and (\ref{excon}) as
reasonable approximations for larger systems, where diagonalization is
impossible.

In order to extract properties, such as ground-state energy and correlations,
and the excitation gap of the system in the thermodynamic limit, we still need
to calculate averages on states (\ref{ansatz}) and (\ref{excon}). For a
$3 \times 3$ system with periodic boundary conditions the constrained Hilbert
space contains 57376 states; however for a $4 \times 4$ system this number is
284 465 424, and regular sums over the configurations are no longer feasible
in a straightforward way. We have thus adopted a Monte Carlo sampling procedure
for this purpose. Energy, for example, is evaluated as the average local energy
$E_{loc}(j)$,
\begin{eqnarray}
E &=& {{\langle \Psi \mid H^{kin\,1} \mid \Psi \rangle} \over
{\langle \Psi \mid \Psi \rangle}} = \left(\langle \Psi \mid \Psi \rangle
\right)
^{-1} \sum_{j} \mid \Psi(j) \mid^2 {{H^{kin\,1} \Psi(j)} \over {\Psi(j)}}
\nonumber \\
&=& \left(\sum_{j} \mid \Psi(j) \mid^2 \right)^{-1}
\sum_{j} \mid \Psi(j) \mid^2 E_{loc}(j) \; , \label{eloc}
\end{eqnarray}
where $j$ labels the configurations of the whole system (and we recall that we
are still discussing the case $J = 0$).

We have calculated the average energies of states (\ref{ansatz}) and
(\ref{excon}) for system sizes up to $30 \times 30$. The relevant information
obtained from this calculation is summarized in Fig.6 and Fig.7. On Fig.6 we
have plotted the energy gap $\Delta E$ versus system size $L$, which shows that
the gap tends to a value close to ${{1}\over{2}}$ as $L \rightarrow \infty$ and
$U \rightarrow \infty$.
Comparison with the unconstrained gap value $\Delta E = 2$ (for $U = J = 0$)
shows that the constraint, even though insufficient to make the paraelectric
state unstable and restore ferroelectricity, does act all the same to increase
very drastically the tendency to ferroelectricity.

On Fig.7 we show a plot of ground state energy per site as a function of $L$,
and we see that the finite size energy corrections scale exponentially with L.
This agrees well with the existence of an excitation gap. The correlation
length obtained from energy corrections is $\xi = 0.72$, which agrees with
absence of long range order. To clarify this point, we calculated the
correlation function $\langle z^{*}_{i} z_{j}\rangle$ in the ground state, and
find that it falls off very quickly with distance.
It was not possible to investigate accurately its large distance behaviour,
because the corresponding values were close to zero and suffered from large
statistical error. For illustration, we present in Table 2 some data for the
$10 \times 10$ system. This behaviour of the correlation function supports the
conjecture drawn from finite-size scaling of the ground state energy, and
leaves
us with very little doubt that ansatz (\ref{ansatz}) for the $J = 0, U =
\infty$
state is disordered.

Unfortunately, we have not been able to construct variational states, yielding
a similar semi-analytical
understanding of the constrained quantum model for a general $J \neq 0$.
In order to get at least a crude prediction for the critical value $J_{q}/t$,
we have done a slave boson mean-field calculation, which is presented in the
Appendix. According to this approximate theory, there is already for
$J = 0, U = \infty$, a
ferroelectric state which is degenerate in energy with the paraelectric one.
This would suggest a critical value $J_{q}/t = 0$, which is not
compatible with the above result obtained from the ansatz (\ref{ansatz}). In
the following chapters, we shall resolve this issue by means of a numerical
Path
Integral Monte Carlo simulation. Before doing so, we present a simple argument
which indicates that $J_{q}/t$ is finite for the constrained 2D quantum
four-state clock model. As it can be easily shown, the dielectric
susceptibility
of a single four-state rotor at zero temperature is $\chi_{0} = 1/2t =
1/\Delta E$,
where $\Delta E$ is the energy gap. Since each rotor has four nearest
neighbours, a simple standard mean field scheme would suggest a para-ferro
transition at $4 \, \chi_{0} \, J_{q} = 1$. For a simple, unconstrained model
this yields a critical value of $J_{q}/t = 1/2$, to be compared with the true
value $0.66$. Since in the constrained case the gap is smaller roughly by a
factor of four, the predicted critical value correspondingly becomes equal to
$J_{q}/t = 1/8 = 0.125$. As we shall see later, this value is again
reasonably accurate. The basic fact that $J_{q}/t$ remains finite, in spite of
the constraint, is thus confirmed.

\section {\bf The Quantum Monte Carlo study}

In this section we shall describe the Path Integral Quantum Monte Carlo
algorithm implemented for the study of our problem. The first problem
to be resolved is the choice of a convenient decomposition of the hamiltonian
in order to apply the Trotter formula.

Since we are interested in the $U \rightarrow \infty$ limit of the hamiltonian
(\ref{ham}), we need to be especially careful. Splitting
naively $H^{tot}$ into the sum of a kinetic and potential energy yields in the
limit $U \rightarrow \infty$ a very poor convergence of the Trotter expansion,
because the commutator of the kinetic energy operator and the unbounded
potential energy operator is also unbounded. The error decreases slowly, as
$1/m$, with the number of Trotter slices $m$, analogously to that of
Ref. \cite{zhang}. In order to avoid this problem, we have to get rid of the
unbounded term in the hamiltonian. For this purpose it is useful to introduce
a projector $P$
\begin{equation}
P = \prod_{\langle ij \rangle} (1 - P_{ij}^{'})
= \prod_{\langle ij \rangle} P_{ij} \; ,
\label{project}
\end{equation}
which projects out the forbidden states. The partition function of
(\ref{ham}) in the limit $U \rightarrow \infty$ then can be written as
\begin{equation}
Z = \mbox{Tr} \,P  e^{-\beta PHP} \; ,
\end{equation}
where we again denote by $H$ the unconstrained hamiltonian (\ref{simple})
\begin{equation}
H = H^{4} + H^{kin\,1} = \sum_{i} ( H^{4}_{i} + H^{kin\,1}_{i} ) =
\sum_{i} H_{i} \; .
\end{equation}

It is now convenient to use both {\it row and column} indices $i,j$ to label
the sites of the square lattice, and write the hamiltonian as
\begin{equation}
H = \sum_{ij} H_{ij} = \sum_{(i+j) \; odd} H_{ij} +
\sum_{(i+j) \; even} H_{ij} =
\sum_{(i+j) \; odd} A_{ij} + \sum_{(i+j) \; even} B_{ij} \, ,
\end{equation}
where $A_{ij}$ is just a re-naming of $H_{ij}$ for the $(i+j)$ odd sublattice,
and $B_{ij}$ that for the even sublattice.
Since both $H$, and $P$, contain only nearest neighbour interactions, we have
\begin{equation}
\left[P A_{ij} P,P A_{i^{'}j^{'}} P \right] =
\left[P B_{ij} P,P B_{i^{'}j^{'}} P \right] = 0 \, , \label{eq:com}
\end{equation}
which provides us with a useful decomposition of the hamiltonian for
application
of the Trotter formula. The whole lattice is decomposed into two
interpenetrating
sublattices, A and B, each sublattice consisting of noninteracting sites in
an external field determined by the configuration of the other sublattice.
This scheme is actually a version of the well-known checkerboard decomposition
\cite{tan}.

To proceed, we write the partition function of the system as
\begin{equation}
Z = \mbox{Tr} P \, e^{-\beta PHP} = \sum_{\{ z_{ij} \}} \langle z_{ij} | P
\left[ e^{- {{\beta}\over{m}} \left( \sum_{ij} PA_{ij}P + \sum_{ij} PB_{ij}P
\right) } \right]^{m} | z_{ij} \rangle \, ,
\end{equation}
where $m$ is an integer, and $| z_{ij} \rangle$ are eigenstates of the complex
coordinates $z_{ij}$ (\ref{complex}).

We define the $m$-th approximant to the partition function by
\begin{eqnarray}
Z_{m} &=& \sum_{\{ z_{ij} \}}
\langle z_{ij} | P \left[ e^{- {{\beta}\over{m}} \sum_{ij} P A_{ij} P}
e^{- {{\beta}\over{m}} \sum_{ij} P B_{ij} P} \right]^{m}
| z_{ij} \rangle \nonumber \\
&=& \sum_{\{ z_{ij}^{1} \}} \ldots \sum_{\{ z_{ij}^{2m} \}}
\langle z_{ij}^{1} | P e^{- {{\beta}\over{m}} \sum_{ij} P A_{ij} P}
| z_{ij}^{2} \rangle
\langle z_{ij}^{2} | P e^{- {{\beta}\over{m}} \sum_{ij} P B_{ij} P}
| z_{ij}^{3} \rangle  \nonumber \\
& & \ldots \langle z_{ij}^{2m-1} |  P e^{- {{\beta}\over{m}} \sum_{ij} P
A_{ij} P}
| z_{ij}^{2m} \rangle \langle z_{ij}^{2m} | P e^{- {{\beta}\over{m}} \sum_{ij}
P B_{ij} P} | z_{ij}^{1} \rangle \, ,
\end{eqnarray}
where we have inserted a complete set of intermediate states between each two
exponentials.
Due to (\ref{eq:com}), the matrix elements between the intermediate states in
the last expression factorize, and we get
\begin{eqnarray}
& &Z_{m} = \sum_{\{ z_{ij}^{1} \}} \ldots
\sum_{\{ z_{ij}^{2m} \}}  \nonumber \\
& &\prod_{k,(i+j) \, odd}
\langle z_{i+1,j}^{k},z_{i,j+1}^{k},z_{i-1,j}^{k},
z_{i,j-1}^{k},z_{i,j}^{k} | P e^{- {{\beta}\over{m}} P A_{ij} P}
| z_{i+1,j}^{k+1},z_{i,j+1}^{k+1},z_{i-1,j}^{k+1},z_{i,j-1}^{k+1},
z_{i,j}^{k+1} \rangle
\nonumber \\
& &\prod_{k,(i+j) \, even} \langle z_{i+1,j}^{k},z_{i,j+1}^{k},z_{i-1,j}^{k},
z_{i,j-1}^{k},z_{i,j}^{k} | P e^{- {{\beta}\over{m}} P B_{ij} P}
| z_{i+1,j}^{k+1},z_{i,j+1}^{k+1},z_{i-1,j}^{k+1},z_{i,j-1}^{k+1},
z_{i,j}^{k+1} \rangle \, ,
\end{eqnarray}
where $| z_{i+1,j}^{k},z_{i,j+1}^{k},z_{i-1,j}^{k},z_{i,j-1}^{k},z_{i,j}^{k}
\rangle = | z_{i+1,j}^{k} \rangle | z_{i,j+1}^{k} \rangle
| z_{i-1,j}^{k} \rangle | z_{i,j-1}^{k} \rangle | z_{i,j}^{k} \rangle $.
It is easily seen that the matrix elements in the last equation are
{\it diagonal} in the quantum numbers
$z_{i+1,j}^{k},z_{i,j+1}^{k},z_{i-1,j}^{k},z_{i,j-1}^{k}$,
and can therefore be written as
\begin{eqnarray}
\langle z_{i+1,j}^{k},z_{i,j+1}^{k},z_{i-1,j}^{k},
z_{i,j-1}^{k},z_{i,j}^{k} | P e^{- {{\beta}\over{m}} P H_{ij} P}
| z_{i+1,j}^{k},z_{i,j+1}^{k},z_{i-1,j}^{k},z_{i,j-1}^{k},
z_{i,j}^{k+1} \rangle \nonumber \\
= C(z_{i+1,j}^{k},z_{i,j+1}^{k},z_{i-1,j}^{k},z_{i,j-1}^{k},z_{i,j}^{k},
z_{i,j}^{k+1}) \, . \label{boltz}
\end{eqnarray}
This diagonality imposes a conservation rule on the 2+1 D classical system on
which our 2D quantum system maps. The mapping now follows from
\begin{eqnarray}
Z_{m} = \sum_{\{ z_{ij}^{1} \}} \ldots
\sum_{\{ z_{ij}^{2m} \}} \prod_{k,i+j \; odd}
C(z_{i+1,j}^{k},z_{i,j+1}^{k},z_{i-1,j}^{k},z_{i,j-1}^{k},z_{i,j}^{k},
z_{i,j}^{k+1}) \nonumber \\
\prod_{k,i+j \; even}
C(z_{i+1,j}^{k},z_{i,j+1}^{k},z_{i-1,j}^{k},z_{i,j-1}^{k},z_{i,j}^{k},
z_{i,j}^{k+1}) \, , \label{eq:map}
\end{eqnarray}
and each term in the above product can be interpreted as a Boltzmann factor
of a parallelepiped with vertical faces rotated by $45^{\circ}$ with respect to
the axes of the original 2D lattice (Fig.8). Each of its corner points is a
four-state clock rotor. The corresponding classical system can thus be
seen as a (2+1) D lattice of such parallelepipeds. The conservation rule
restricts {\it the rotors on both sides of the vertical
edges of each parallelepiped to be in the same state}. The Boltzmann weight
(\ref{boltz}) of each parallelepiped then depends on the state of all four
rotors at its vertical edges, as well as on the state of both rotors at the
centers of its horizontal faces. The mutual state of these latter
rotors, sitting at the centers of the horizontal faces, {\it is not restricted}
in any way.

The calculation of the matrix elements $C(z_{i+1,j}^{k},z_{i,j+1}^{k},
z_{i-1,j}^{k},z_{i,j-1}^{k},z_{i,j}^{k},z_{i,j}^{k+1})$ is straightforward.
It requires just diagonalization of the $4 \times 4$ on-site problem of a
four-state clock rotor $H^{kin\;1}_{i}$ in an external field, represented by
$H^{4}_{i}$ and the projector $P$. In our simulation we performed this
diagonalization numerically. All the matrix elements turned out to be
non-negative, and therefore no sign problem was present \cite{foot2}.

We proceed to identify the estimators for the thermodynamic quantities.
For convenience we write the partition function approximant $Z_{m}$
symbolically as
\begin{equation}
Z_{m} = \sum_{\{ z_{ij}^{1} \}} \ldots
\sum_{\{ z_{ij}^{2m} \}} \prod_{k,i,j} C_{kij} \, .
\end{equation}
The internal energy per site of the system then reads
\begin{equation}
E_{m} = - {{1} \over {N}} {{1} \over {Z_{m}}}
{{\partial Z_{m}} \over {\partial \beta}} = \langle E_{est} \rangle \, ,
\end{equation}
where
\begin{equation}
E_{est} = - {{1} \over {N}} \sum_{k,i,j} {{1} \over {C_{kij}}}
{{\partial C_{kij}} \over {\partial \beta}}
\end{equation}
is the corresponding estimator.
For the specific heat per site $c_{v}$ we obtain
\begin{eqnarray}
c_{vm} &=& - {{1} \over {N}} \beta^{2}
{{\partial E_{m}} \over {\partial \beta}} \nonumber \\
&=& {{1} \over {N}} \beta^{2}
\left\{ \langle E_{est}^{2} \rangle - \langle E_{est} \rangle^{2} +
\langle \sum_{k,i,j} \left[ {{1} \over {C_{kij}}}
{{\partial^{2} C_{kij}} \over {\partial \beta^{2} }}
- \left( { {1} \over {C_{kij}} } {{\partial C_{kij}} \over {\partial \beta}}
\right)^{2} \right] \rangle \right\} \, ,
\end{eqnarray}
and we see that it is not just equal to the fluctuation of the internal
energy, as in the classical case. The hamiltonian of our effective
classical system is itself temperature dependent.

In order to calculate the order parameter, which in our case is the
polarization
of the system, and the corresponding susceptibility, we have to consider an
external field $F$ applied on the system. If we choose this to point, e.g.
along the $x$-axis, the matrix elements $C(z_{i+1,j}^{k},z_{i,j+1}^{k},
z_{i-1,j}^{k},z_{i,j-1}^{k},z_{i,j}^{k},z_{i,j}^{k+1})$ in (\ref{eq:map}) are
replaced by
\begin{eqnarray}
& &C(z_{i+1,j}^{k},z_{i,j+1}^{k},
z_{i-1,j}^{k},z_{i,j-1}^{k},z_{i,j}^{k},z_{i,j}^{k+1}) \rightarrow \nonumber \\
& &C(z_{i+1,j}^{k},z_{i,j+1}^{k},z_{i-1,j}^{k},z_{i,j-1}^{k},z_{i,j}^{k},
z_{i,j}^{k+1}) \exp( {{\beta}\over{m}} F \mbox{Re} \, z_{i,j}^{k} )
\end{eqnarray}
and we obtain for the equilibrium polarization per site $P_{xm}$
\begin{equation}
P_{xm} = {{1} \over {\beta N Z_{m}}} {{\partial Z_{m}} \over {\partial F}} =
\langle P_{xm \, est} \rangle \, ,
\end{equation}
where
\begin{equation}
P_{xm \, est} = {{1} \over {N m}} \sum_{k,i,j} \mbox{Re} \, z_{i,j}^{k}
\end{equation}
The static dielectric susceptibility $\chi_{xx \, m}$ then turns out to be
\begin{equation}
\chi_{xx \, m} = {{\partial P_{xm}} \over {\partial F}} =
\beta N \left[ \langle P_{xm \, est}^{2} \rangle - \langle P_{xm \, est}
\rangle^{2} \right] \, .
\end{equation}
The last two quantities are suitable to study the system either in the
paraelectric
phase, where $\langle P_{xm \, est} \rangle = 0$, or in the ferroelectric phase
not very close to the transition, when the order parameter does not undergo the
finite-size flipping among the four possible orientations, and
$\langle P_{xm \, est} \rangle$ remains a well-defined quantity during the
simulation time. For the
behaviour of the system right across the phase transition, it is convenient to
monitor the modulus rather than the components of the order parameter. We have
\begin{equation}
P_{m} = \langle P_{m \, est} \rangle = \langle \sqrt{P_{xm \, est}^{2} +
P_{xm \, est}^{2}} \rangle \, ,
\end{equation}
and the corresponding susceptibility is given by
\begin{equation}
\chi_{m} = \beta N \left[ \langle P_{m \, est}^{2} \rangle -
\langle P_{m \, est} \rangle^{2} \right]\,.
\end{equation}
Finally, we have sampled a "short-range order parameter"
$\langle \cos(\phi_{ij} - \phi_{i^{'}j^{'}}) \rangle$, where $ij$ and
$i^{'}j^{'}$ are nearest neighbours. This can be calculated as
\begin{equation}
\langle \cos(\phi_{ij} - \phi_{i^{'}j^{'}}) \rangle = {{1} \over {2 N m}}
\langle \sum_{k,i,j} \left[ \mbox{Re} ({z_{i,j}^{k}}^{*} z_{i+1,j}^{k}) +
\mbox{Re} ({z_{i,j}^{k}}^{*} z_{i,j+1}^{k}) \right] \rangle \, .
\end{equation}

Our (2+1) dimensional classical system is simulated in a standard way using the
Metropolis algorithm. The trace operation requires periodic boundary conditions
along the imaginary time direction. We have used periodic boundary conditions
for both space directions as well. In order to satisfy the conservation rule,
we must always move
the pairs of rotors on both sides of the vertical edges of the parallelepipeds
simultaneously. It is therefore convenient to consider the vertical edges as
being a kind of "rigid rods", and take these as new variables, which now can
be moved independently (Fig.8). One randomly chosen rod was moved at a time,
without any kind of collective moves. Unlike in simulation of the classical
four-state clock model \cite{tobochnik}, we did not restrict the flips of the
rods and allowed these to flip from present position to each of the remaining
three.

We carried out almost all calculations by fixing the value of the ferroelectric
coupling $J$ and of the number of Trotter slices $m$, and then running a series
of simulations for different temperatures. The final configuration of a
simulation
at a given temperature was used as the initial configuration for a run at the
next higher temperature, always heating the system. As the initial
configuration
at the lowest temperature, we always took the ferroelectric state, completely
ordered both in space and in the imaginary time directions. The only exception
is the data shown on Fig.18, where the corresponding runs were carried out at
constant temperature, decreasing the value of $J$. We typically used
$1 \div 2 \times 10^{4}$ MCsteps/site to equilibrate the system and
$2 \div 4 \times 10^{5}$ MCsteps/site for calculating averages.
The CPU time needed to perform $1$ MCstep/site was about $14 \, \mu s$ using an
HP720 RISC machine.

In order to estimate the statistical accuracy of our results we measured the
MC correlation time of the chain of values generated for various
quantities in course of the simulation. We did this by using the standard
method
of dividing the chain in blocks of variable size described in \cite{jaccucci},
\cite{caoberne}.

Before describing our results in the two following sections, we wish to comment
briefly on the convergence of the averaged quantities as a function of the
number of Trotter slices $m$. It is well-known \cite{tan} that the
error in the average value of an operator due to the Trotter decomposition is
an even function of $m$ and therefore the results of a QMC simulation can, in
principle, be extrapolated to $m \rightarrow \infty$ in the form
\begin{equation}
A_{m} = A_{\infty} + {{a} \over {m^{2}}} + {{b} \over {m^{4}}} + \ldots \, .
\end{equation}
This kind of extrapolation, to be meaningful, would require very high accuracy
for at least two values of $m$ (to determine $a$), or three values of $m$ (to
determine $b$), etc. For most data presented in this paper such
procedure was not needed. Except at the lowest
temperatures, it was generally sufficient to inspect data for $m = 5, m = 10$
or $m = 10, m = 20$ (sometimes for all three values of $m$) to see that further
increasing of $m$ would not change the value of the quantity under
consideration
within our statistical accuracy. The values of $m$ we have
used were always a result of a compromise between the requirement of
convergence
and that of keeping a reasonable acceptance ratio, since, of course, the
acceptance ratio falls with increasing $m$ and problems with
loss of ergodicity of the system begin to appear.

\section {\bf Test of the PIMC scheme: the unconstrained four-state clock
model.}

In order to test the QMC scheme described in the last section, we chose to
perform a simulation of a limited extent for the simple unconstrained model,
which is equivalent to the quantum Ising model, as discussed in section 3.
In this section we briefly describe the results, since they will also be of
interest for comparison with those obtained for the constrained model in the
next section.

{}From the Monte Carlo renormalization group study of Kolb \cite{kolb} it is
known
that the limiting value of the ratio $t/J$ for the existence of an ordered
state
in 2D Ising model in transverse field is $3.04$. Using the mapping described in
section 3 then results in a limiting value of $J_{q}/t = 0.66$ for our model.

As an initial test, we did a numerical diagonalization of a $2 \times 2$ system
with $J = 0.5$, and compared the exact canonical results for internal energy
and specific heat with those of a PIMC simulation.
In order to obtain a good statistical accuracy, we used $1 \times 10^{6}$
MCsteps/site for the simulation, and found excellent agreement (Fig.9).

Next, we carried out simulations for three different
values of $J = 0.5, 0.75, 1.0$. Each of these three values of $J$ is
expected to correspond to a different regime of the system. The results can be
described as follows.

We start with $J = 1.0$, where we expect the system to approach the
classical region of the phase diagram. It should therefore exhibit onset of
ferroelectricity with a fully developed critical behaviour of the 2D Ising
universality class, already for the relatively small system sizes we studied.
The results of the simulation performed for lattice sizes $L=6,10,20$ are
plotted on Fig.10.
We see that the zero temperature value of the order parameter $| P |$ is
about $0.85$, only moderately depressed from its classical value by quantum
fluctuations. The transition is signaled by a drop of the order parameter as
well as by pronounced critical peaks of both the dielectric susceptibility
and the specific heat. The susceptibility peak appears at a slightly higher
temperature than the specific heat peak, and is also considerably more size
dependent. All these results are perfectly compatible with a 2D Ising
transition.
In order to extrapolate the infinite-size critical temperature we have analyzed
our data using the phenomenological renormalization method \cite{barber}, which
treats properly the large finite-size corrections (especially evident for the
specific heat peak position). The results of this analysis are given on Fig.11
and Fig.12. In the case of susceptibility, Fig.11,
the three curves are near intersection in the vicinity of the
point $(1.,1.75)$, which suggests the critical temperature $T_{c} = 1.0$ and
the 2D Ising value of ${{7} \over {4}}$ for the ratio of critical exponents
${{\gamma} \over {\nu}}$. The same analysis for the specific heat, Fig.12,
however, suggests instead a higher critical temperature close to $T_{c} = 1.2$.
We consider this discrepancy to be a consequence of small system sizes used as
well as of the statistical errors in the results. In any case, a critical
temperature $T_{c} \sim 1.0$ is rather close to its classical value
$T_{c} = 1.1346 \ldots$, which further confirms the conclusion that the system
for $J = 1.0$ is in a classical regime.

For $J = 0.75$ the simulation was also done for three different lattice sizes
$L=6,10,20$ and the corresponding results are on Fig.13. The zero temperature
value of $| P |$ is now about $0.65$, which reveals the effect of stronger
quantum fluctuations. The ferroelectric-paraelectric transition is seen as a
drop of $| P |$ and as a sharp critical peak of the dielectric
susceptibility curve near a new $T_{c} \sim 0.5$. The same finite size scaling
analysis of the susceptibility as in the preceding case (Fig.14) now agrees
with
a critical temperature of $T_{c} \sim 0.5$ and a ratio of critical exponents
${{\gamma} \over {\nu}}$ again close to ${{7} \over {4}}$. Since the
classical transition temperature for $J = 0.75$ is $T_{c} = 0.851$, the effect
of quantum fluctuations has been to reduce $T_{c}$ very considerably. An
interesting feature in this case is the behaviour of the specific heat. This
has a rather flat maximum at temperatures almost twice as high as the
transition
temperature, with no strong evidence of critical behaviour at $T_{c}$ itself,
for the system sizes we studied. This is again a sign that the system is
in a quantum regime, and one would have to go to larger sizes to observe
the expected crossover \cite{pfeuty1} from quantum to classical critical
behaviour, sufficiently close to $T_{c}$.

Finally, for $J = 0.5$, we ran the simulation for only one lattice size, namely
$L = 10$. The results are on Fig.15.  Ferroelectricity seems altogether
absent, all the way down to $T = 0$.
To clarify further this case, we also performed an
extrapolation of the dielectric susceptibility to $m \rightarrow \infty$ and
found it to saturate at low temperatures (for comparison with the classical
case, the susceptibility starts to level off at temperature $T \sim 0.25$,
which
is slightly less then a half of the classical transition temperature
$T_{c} = 0.567$). This means that the paraelectric state persists down to the
lowest temperatures and we therefore conclude that for this value of $J$ the
system is in the quantum paraelectric regime.

Apart from the usual long-range order parameter $| P |$, we can also
demonstrate
the temperature dependence of the nearest-neighbour "short-range order
parameter" $\langle \cos(\phi_{ij} - \phi_{i^{'}j^{'}}) \rangle$ [Fig.16].
Since the system is now paraelectric for all temperatures, $| P |$ must
scale to zero with increasing system size L. Its behaviour for a finite L
reflects that of the correlation length $\xi$. The polarization $|P|$ is seen
to pass through a moderate {\it maximum} at temperature $T^{*} \sim 0.6$, where
the same kind of behaviour is clearly visible also on the nearest-neighbour
order $\langle\cos(\phi_{ij} - \phi_{i^{'}j^{'}})\rangle$ curve. The
correlation
length $\xi$ of a system in the quantum paraelectric regime thus has a maximum
at a finite $T^{*}$. This effect, in the different context of granular
superconductors, was found by Fazekas et al. in \cite{faz}, where a continuous
XY-model was investigated. Their conclusion as to the existence of a broad
maximum of short range order at $T^{*} \sim J$ remains therefore valid also in
the case of our discrete four-state clock model.
Qualitatively, the interpretation of this effect is the following.
At zero temperature, the rotors are predominantly in their totally symmetric
ground state, corresponding to the angular momentum $j = 0$, which does not
possess a dipole moment. Increasing temperature from $T=0$, rotor states with
non-zero dipole moment become thermally excited, and the system starts to
develop some kind of short range order due to the coupling $J$. This order
reaches a maximum at finite temperature, and is then disrupted by thermal
fluctuations as temperature increases further.

\section{Simulation of the constrained quantum four-state clock model.}

Before coming to the actual results of the simulation of the constrained
model, we must again recall the limitations of the method used. It is
intuitively clear that already in the classical case, the presence of the
constraint acts to reduce very considerably the acceptance ratio of our simple
unbiased MC moves. Moreover, in
the quantum case, the constraint increases the systematic error due to Trotter
decomposition, and a larger number of Trotter slices is needed to approach the
true quantum averages. These problems seriously reduce our possibilities of
performing a satisfactory finite-size scaling analysis of the results
\cite{foot3}.

As a test of the code, we simulated first a $2 \times 2$ constrained four-state
clock model, for $J=0.5$, and compared the results with an exact
diagonalization, with very satisfactory agreement (Fig.17). The rest of results
we present in this section refers to a single system size, $L=10$. The
simulation
was performed typically for $m=10$ Trotter slices in the temperature interval
from $0.1 < T < 1$. In the low temperature range ($T \leq 0.4$) we also used
$m=20$ and $m=40$ Trotter slices. All the quantities were averaged over
$4\times10^5$ MCsteps/site, after discarding the initial
$2\times10^4$ MCsteps/site necessary for equilibration.

Unlike the unconstrained clock model, in the constrained case we lack accurate
predictions for the quantum critical value $J_{q}/t$, where a $T=0$ quantum
ferro-para transition takes place. The variational ansatz (\ref{ansatz})
suggests that $J_{q} > 0$. On the other hand, we know, that the constraint can
only drive the system closer to ferroelectricity, which implies
$J_{q}/t < J_{q}^{0}/t = 0.66$. In Fig.18, we plot the calculated constrained
dielectric susceptibility $\chi$ versus coupling constant $J$ at
low temperatures, $T=0.1$ and $T=0.2$. We see a sharp peak, signalling
a transition, roughly at $J_{c} \sim 0.17$ for $T=0.2$ and at $J_{c} \sim 0.15$
for $T=0.1$. The limiting quantum critical value may thus be estimated to be
$J_{q}/t \sim 0.15$, a factor of about four smaller than that
for the unconstrained model.

On Fig.19 we plot the order parameter $|P|$, the dielectric susceptibility
$\chi$ and the specific heat $c_{v}$ versus temperature for $J = 0.3$. The
system is clearly ferroelectric at low temperatures. However, the saturation
value of the order parameter $|P|$ is $\sim 0.5$, indicating a strong quantum
fluctuation reduction. The susceptibility $\chi$ has a peak at $T \sim 0.35$,
for this value of $J$. At the same temperature, however, the specific heat
$c_{v}$ does not show any apparent singularity for this small size. As
confirmed
also by results of the previous section, milder specific heat singularities
are,
however, quite typical for quantum transitions \cite{opp},\cite{morf}.

Fig.20 and Fig.21 present plots of the same quantities $\chi_{xx}$, $c_{v}$ and
$|P|$, plus in addition the temperature dependence of the nearest-neighbour
short range order parameter $\langle\cos(\phi_{ij} -
\phi_{i^{'}j^{'}})\rangle$,
for $J=0.05$. For this value of the coupling constant, which is lower than
$J_{q}/t$, the system should be in the quantum paraelectric regime at low $T$.
Actually, were it not for the $J$ dependence of $\chi$ (Fig.18), it would
be difficult to draw this conclusion just by inspection of the $T$-dependence
of
$\chi_{xx}$. In fact, even at the lowest temperatures investigated, $\chi_{xx}$
does not seem to saturate, and continues to grow with decreasing $T$.
Further simulation for even lower temperatures in this deeply quantum regime is
at the moment not feasible, since the Trotter error would be too large.

Similarly to the unconstrained case, the polarization $|P|$ is seen to pass
through a moderate maximum at temperature $T^{*} \sim 0.3$, and the same
behaviour is now just barely visible also on the nearest-neighbour order
$\langle \cos(\phi_{ij} - \phi_{i^{'}j^{'}}) \rangle$. Lastly, we note, that
the specific heat $c_{v}$ is smooth, with the same broad maximum near
$T\sim 0.8$ present also for the ferroelectric case $J = 0.3$. Finally, on
Fig.22, we show a sketch of phase diagrams for both the unconstrained and the
constrained model, as resulting from our simulations.

\section{Discussion, and Conclusions.}

In this paper, we have studied the physics of the simplest model quantum
paraelectric, both as a function of coupling at zero temperature, and at
finite temperatures. The selection of a meaningful hamiltonian is particularly
delicate, and we have been able to make only some tentative steps towards
modeling the real perovskite quantum paraelectrics $SrTiO_{3}$ and $KTaO_{3}$.
Three main points have emerged in this selection process.

The first is that only models capable of describing order-disorder fluctuations
should be considered adequate, since experimentally quantum perovskites appear
to crossover into an order-disorder regime, before entering a sort of quantum
central peak state at low temperatures. An
alternative displacive, mean-field picture would be inadequate to describe this
situation. We have therefore preferred a lattice model, and chosen to work
exclusively with techniques, which can describe adequately fluctuations,
quantum
as well as classical. The basic element in our lattice models is the $Ti - O$
dipole, also called a "bond".

A second point, which has emerged, is that we might expect in a perovskite
ferroelectric at least two different quantum tunneling processes. The first is
bond tunneling between the $n$ equivalent positions inside the octahedral cage
($n=4$ in tetragonal $SrTiO_{3}$, $n=6$ in cubic $KTaO_{3}$). The second is
bond tunneling between one cage and the next. This second process will occur
whenever an oxygen which is bonded to one $Ti$ hops to bond the other $Ti$, as
in the Fig.2. In this first study, however, we have included only the first
processes.

A third important point is that there are two strong constraints which restrict
both classical and quantum mechanical motion, if the model is to make physical
sense for real perovskites. A first constraint, probably valid at low
temperatures, is that there should not be more than one bond per cell. In other
words, so long as we are in the order-disorder regime, and all $Ti$ atoms are
instantaneously off-center, each can engage one and only one of the surrounding
oxygens. The second, more interesting
constraint, is that two neighbouring cages must not simultaneously possess
bonds which point towards one another. This constraint comes from the physical
impossibility of an oxygen to be engaged in two bonds simultaneously, and
constitutes a kind of ice-rule. The necessity of an effective ice-rule in
displacive perovskite ferroelectrics (or, for that matter, in cuprate high
$T_{c}$ superconductors !) does not seem to have been noted and made use of
before, and might have far-reaching consequences.

Guided by these considerations, we have selected a short-range, two-dimensional
lattice four-state quantum clock model, as the simplest toy model of quantum
paraelectrics. Of the ingredients mentioned above, two in particular pose some
difficulty, namely the inter-cell bond tunneling, and the ice-rule constraint.
Our strategy has been therefore to start off first without these two
ingredients, with the simple quantum four-state clock model, and then to add
the
complications only gradually and successively. Since handling of bond hopping
is
very dependent from the presence of the ice constraint, it was necessary to
include the latter first. Thus, the second model considered in this paper is an
ice-rule constrained quantum four-state clock model. This still leaves out bond
hopping, whose inclusion is now being actively considered, and will hopefully
form the subject of the subsequent paper.

The main method adopted for the present study of the unconstrained and
constrained 2D quantum four-state clock model is Path Integral Monte Carlo,
PIMC, plus finite size scaling when possible. For the unconstrained case this
calculation is meant to reproduce known results, since, as we show, the model
maps onto two uncoupled quantum Ising models (Ising model in a transverse
field), well studied and characterized in the 70's. The ice-rule constrained
model is instead new and non-trivial. A Monte Carlo strategy has been devised,
in order to eliminate a pathologically slow $1/m$ convergence with the number
of
time slices $m$, to recover even for the constrained model the usual, more
comfortable $1/m^{2}$ convergence. Although we cannot push finite size scaling
to produce critical exponents for this case, we have obtained a reasonably
accurate phase diagram.

There are, both in the constrained and in the unconstrained model, only two
phases, the ferroelectric and the paraelectric states. However, the ice-rule
constraint greatly reduces the number of excited configurations, and this in
turn reinforces ferroelectricity. As a result, the extrapolated $T=0$ critical
ratio of potential to kinetic coupling parameters is $J_{q}/t \simeq 0.15$,
a factor
of about four smaller than that $J_{q}^{0}/t \simeq 0.66$ of the unconstrained
model. The classical critical temperature will also be raised with respect to
the unconstrained case, although we did not pursue this aspect in detail. We
would not be surprised if a future closer analysis of both classical and
quantum
critical behaviour should reveal different universality classes, for the
unconstrained and the ice-rule constrained 2D clock models.

What have we learned on the nature of the QPE state in the constrained and
unconstrained quantum four-state clock model ? We have found that the state is
fully symmetric, and has an excitation gap in both cases. However, the gap
$\Delta E$ is very severely reduced by the ice-rule constraint. This reduction
is at least a factor of four at $J=0$, but seems even stronger for finite
$J$. So long as the ground state is nondegenerate with an excitation gap,
transformation from the classical paraelectric state at high temperatures to
the
QPE state at low temperatures is predicted to be a smooth crossover, not a
phase
transition. The finding is in line with traditional views on QPE's \cite{mul2}.
It does not explain, however, the phase transition phenomena found by
M\"{u}ller
et al. in pure, tetragonal $SrTiO_{3}$, and also those seen in low
concentration
uniaxial $KTa_{1-x}Nb_{x}O_{3}$, even at low $x$, well below the onset of
long-range ferroelectric order \cite{rytz}.

To put things in the right perspective, we should however remember, that the
basic assumption behind our lattice model is, as stressed in the introduction,
that the dipoles, i.e. stronger $Ti - O$ or $Ta - O$ bonds, should exist at all
temperatures. In reality, this is not the case. NMR data of Rod, Borsa and
van der Klink \cite{rod} clearly indicate that off-center displacement of $Ta$
ions sets up rather abruptly below $T^{*} \sim 40 K$. A very similar picture
is probably valid also for $SrTiO_{3}$. In other words, crossover from the
high-temperature displacive regime, with a soft mode and no distortion, and the
order-disorder regime occurs suddenly, and very close to the extrapolated
displacive Curie temperature $T^{*}$. The individual dipoles of our discrete
model are born as such at this abrupt crossover temperature. The model itself,
therefore, makes sense only for $T < T^{*}$. What should turn the expected
sharp
crossover into a genuine phase transition in $SrTiO_{3}$, remains at this stage
unclear.

We are presently pursuing further the physics of these interesting phenomena,
by including next the bond-hopping term, and hope to report relevant results in
the near future.

\nonum
\section{Acknowledgements}
We wish to acknowledge fruitful discussions with, and help from, E. Courtens,
A. Ferrante, M. Maglione, G. Mazzeo, K. A. M\"{u}ller and S. Sorella.
Support by INFM and by the European Research Office, U.S. Army, is also
acknowledged.

\nonum
\section{Appendix: a slave-boson mean field theory}

In this appendix we present a mean field theory for the ground state of the
constrained model, at $J = 0$. From this, we also get an estimate for the
critical value $J_{q}/t$. The main problem is to deal with the rigid constraint
which prevents the oxygens from being doubly bonded. As mentioned, this feature
of our model
is very similar to that which applies to electrons in the infinite U Hubbard
model, where the repulsion energy also eliminates double occupancy of a single
site. This analogy suggests that we may try to use the method of introducing
auxiliary, slave-boson fields, commonly used in the Hubbard model studies
\cite{rucken}.

For this purpose it is convenient to work in the occupation number
representation. First of all, for each pair of nearest neighbour sites $i,j$ we
introduce two Bose operators $b^{\dagger}_{ij}, b^{\dagger}_{ji}$. The operator
$b^{\dagger}_{ij}$ creates a bond of the central ion on site $i$, pointing
towards site $j$, and similarly $b^{\dagger}_{ji}$ creates a bond of the
central ion on site $j$ pointing towards site $i$. We notice that obviously
$b^{\dagger}_{ij} \neq b^{\dagger}_{ji}$. Between the central sites $i,j$,
there
is a "bridging" oxygen ion, which can be unambiguously labeled by the pair of
site labels $i,j$ or $j,i$. This oxygen ion, depending on the states of central
ions on sites $i,j$, can in fact be in four different states. In order to
describe these states, we introduce new Bose operators $e^{\dagger}_{ij},
p^{\dagger}_{ij,i},p^{\dagger}_{ij,j},d^{\dagger}_{ij}$, with the following
meaning. The oxygen is in the state $e^{\dagger}_{ij} \, | 0\rangle $, if
none of the central ions is bonded to it. It is in the state
$p^{\dagger}_{ij,i} \, | 0\rangle $, if the ion on site $i$
is bonded to it, and vice versa for $p^{\dagger}_{ij,j} \, | 0\rangle $.
Finally, in the state $d^{\dagger}_{ij} \, | 0\rangle $, both central ions on
sites $i,j$ are bonded to it. These oxygen operators represent our auxiliary
fields, or slave bosons. In our case they do
correspond to real physical states of the oxygen ion. From this point of view,
they are in fact real bosons; but they are slaved by the hamiltonian itself.
{}From the way we introduced all the operators it is clear that these satisfy
the
following constraints:
\begin{eqnarray}
& & \sum_{j} b^{\dagger}_{ij} b_{ij} = 1 \label{slaveb1}\\
& & e^{\dagger}_{ij} e_{ij} + p^{\dagger}_{ij,i} p_{ij,i} +
p^{\dagger}_{ij,j} p_{ij,j} + d^{\dagger}_{ij} d_{ij} = 1 \label{slaveb2}\\
& & b^{\dagger}_{ij} b_{ij} = p^{\dagger}_{ij,i} p_{ij,i} +
d^{\dagger}_{ij} d_{ij} \label{slaveb3}
\end{eqnarray}

Now we express our hamiltonian (\ref{ham}) by means of the new
operators. First, the constraint term (\ref{constr}) is now easily written as
\begin{equation}
H^{constr} = \lim_{U \rightarrow \infty }
U \sum_{\langle ij \rangle} d^{\dagger}_{ij} d_{ij} \; .
\end{equation}
The potential energy term is straightforwardly written as
\begin{equation}
H^{4} = -J \sum_{\langle ij \rangle} \sum_{kl} \mbox{Re} (r_{ik} r^{*}_{jl})
b^{\dagger}_{ik} b_{ik} b^{\dagger}_{jl} b_{jl}
\end{equation}
The hopping term (\ref{disc}) is slightly more involved. We work in an
enlarged Hilbert space and the hopping of central ion is accompanied by a
change of state of the surrounding oxygens. The corresponding expression reads
\begin{equation}
H^{kin\;1} = -t \sum_{i} \sum_{ jj^{'} }
b^{\dagger}_{ij} ( d^{\dagger}_{ij} p_{ij,j} + p^{\dagger}_{ij,i} e_{ij} )
b_{ij^{'}} ( p^{\dagger}_{ij^{'},j^{'}} d_{ij^{'}} + e^{\dagger}_{ij^{'}}
p_{ij^{'},i} ) \; ,
\end{equation}
where $j,j^{'}$ are both nearest neighbours of $i$ and such that $j^{'}$ is the
next-nearest neighbour of $j$.

Next, we shall treat the {\it oxygen} operators in the mean-field
approximation,
replacing them by $c$-numbers. Before passing to the actual constrained
case $U = \infty$, we consider the trivial case of $U = 0$, $J = 0$.
The corresponding operator averages are obviously equal for all oxygens and
we denote them as
\begin{eqnarray}
& & \langle e^{\dagger}_{ij} \rangle = e \label{av1}\\
& & \langle p^{\dagger}_{ij,i} \rangle = \langle p^{\dagger}_{ij,j} \rangle = p
\label{av2} \\
& & \langle d^{\dagger}_{ij} \rangle = d \label{av3} \; ,
\end{eqnarray}
where the equation (\ref{av2}) expresses the fact that we assume an unbroken
symmetry case.
The constraints (\ref{slaveb1}),(\ref{slaveb2}),(\ref{slaveb3}) then read
\begin{eqnarray}
& & \sum_{j} \langle b^{\dagger}_{ij} b_{ij} \rangle = 1 \nonumber\\
& & e^{2} + 2 p^{2} + d^{2} = 1 \nonumber\\
& & \langle b^{\dagger}_{ij} b_{ij} \rangle = p^{2} + d^{2}
\label{slavebmf}
\end{eqnarray}
Since for $U = 0$ the sites are independent, the numerical values of the above
averages follow just from the statistical distribution of possible mutual
orientations of clock variables on neighbouring sites. It is easily found that
in this case $\langle b^{\dagger}_{ij} b_{ij} \rangle = {{1}\over{4}}$, and
$d^{2} = {{1}\over{16}}$, $p^{2} = {{3}\over{16}}$, $e^{2} = {{9}\over{16}}$.
Our mean-field hamiltonian then becomes a sum of on-site terms
\begin{equation}
H^{MF} = - \tilde{t} \sum_{i} \sum_{ jj^{'} }
b^{\dagger}_{ij} b_{ij^{'}} \; , \label{mf}
\end{equation}
where $\tilde{t} = t p^{2} (d + e)^{2} = {{3}\over{16}} t$. We see that in
order to recover the original $t$ we have to renormalize the mean-field
$\tilde{t}$ by a factor of ${{16}\over{3}}$.

Now we pass to $U = \infty, J = 0$. Obviously, the presence of the constraint
$H^{constr}$ amounts to setting $d = 0$. We shall first search for an unbroken
symmetry, paraelectric ground state. Following the same line of arguments as
above, we have
$p^{2} = {{1}\over{4}}$, $e^{2} = {{1}\over{2}}$, and we get a renormalized
$\tilde{t} = {{2}\over{3}} t$. This corresponds to a ground state energy per
site equal to $E_{g}/N = -2 \tilde{t} = - {{4}\over{3}} t = -1.333 t$, which
compares very well with the result $-1.3668 t$ obtained from our ground state
wavefunction ansatz (\ref{ansatz}) (Fig.7).

Now we show that for $U = \infty, J = 0$ there is also a broken symmetry,
ferroelectric mean-field ground state, whose energy is degenerate with the
paraelectric state found above. First of all we notice that the oxygens
actually form two interpenetrating square sublattices -- one formed by oxygens
lying on horizontal links and other formed by those on vertical links. If we
want to search for a broken symmetry solution, we have to allow for different
oxygen operator averages on these two sublattices, and also the $p$ defining
equation (\ref{av2}) may not be true anymore. Let us assume that the
symmetry is broken along the horizontal axis. Then we shall have nonzero
averages $e_{V}$,$p_{V}$, defined by (\ref{av1}),(\ref{av2}) for oxygens on
links $i,j$, $j=i+\vec{y}$, and $e_{H} = \langle e^{\dagger}_{ij} \rangle$,
$p_{H+} = \langle p^{\dagger}_{ij,i} \rangle$ and
$p_{H-} = \langle p^{\dagger}_{ij,j} \rangle$ for $i,j$, $j=i+\vec{x}$, where
$\vec{x},\vec{y}$ are unit lattice vectors. As a consequence, the mean-field
constraints (\ref{slavebmf}) will also be correspondingly generalized.
Instead of (\ref{mf}) we have now
\begin{equation}
H^{MF} = - \sum_{i} \left\{ \tilde{t}_{1}
(b^{\dagger}_{i,i+\vec{y}} b_{i,i+\vec{x}} +
b^{\dagger}_{i,i+\vec{x}} b_{i,i-\vec{y}}) +
\tilde{t}_{2} (b^{\dagger}_{i,i-\vec{x}} b_{i,i+\vec{y}} +
b^{\dagger}_{i,i-\vec{y}} b_{i,i-\vec{x}}) + \mbox{h.c.} \right\}
\;,\label{mfp}
\end{equation}
where $\tilde{t}_{1} = t \, p_{V}\, e_{V}\, e_{H}\, p_{H+}$,
$\tilde{t}_{2} = t\, p_{V}\, e_{V}\, e_{H}\, p_{H-}$.
We can now prove that $e_{V} = e_{H} = p_{H+} = {{1}\over{\sqrt{2}}}$,
$p_{V} = {{1}\over{2}}$, $p_{H-} = 0$, is a selfconsistent ground state
solution
of (\ref{mfp}). We get renormalized hopping parameters
$\tilde{t}_{1} = {{16}\over{3}} t/4 \sqrt{2}$, $\tilde{t}_{2} = 0$, and
the corresponding normalized ground state of (\ref{mfp}) is
\begin{equation}
| \Psi_{g} \rangle = \prod_{i} \left(
{{1}\over{\sqrt{2}}} b^{\dagger}_{i,i+\vec{x}} +
{{1}\over{2}} b^{\dagger}_{i,i+\vec{y}} +
{{1}\over{2}} b^{\dagger}_{i,i-\vec{y}} \right) | 0 \rangle \; . \label{gs}
\end{equation}
The self-consistency condition is easily found to be satisfied, and the
solution
is clearly ferroelectric.
The corresponding energy per site is $E_{g}/N = - \sqrt{2} \tilde{t}_{1} =
- {{4}\over{3}} t$, and therefore this state is degenerate with the above found
paraelectric state. Our mean-field theory thus predicts a critical value of
$J_{q}/t = 0$.

\figure{Intra-cage bond hopping, corresponding to kinetic energy
$H^{kin\,1}$ (\ref{disc}). }
\figure{Inter-cage bond hopping, due to oxygen tunneling, corresponding to
kinetic energy $H^{kin\,2a}$ (\ref{kin2a}), and $H^{kin\,2b}$ (\ref{kin2b}). }
\figure{The simplest concerted bond hopping mechanism, taking place on the
elementary plaquette. }
\figure{Forbidden configuration, with two bonds sharing the same bridging
oxygen.}
\figure{1D model chain of cages with constraint on each other link.}
\figure{Size dependence of the variational energy gap $\Delta E$ for
$J=0,\;U=\infty$, between ground state (\ref{ansatz}) and excited state
(\ref{excon}).}
\figure{Finite-size scaling to the variational ground state energy
$E_{g}^{\infty}(L)$ for \\ $J=0,\;U=\infty$. Note the clear exponential
behaviour, compatible with a gap, as in the Fig.6.}
\figure{The (2+1)D classical system. The imaginary time (Trotter) direction is
vertical. Vertical dark lines connect sites, which must be flipped together,
springs connect independent sites.}
\figure{Test of the PIMC method: specific heat $c_{v}$, and internal energy
$E$ for a $2 \times 2$ unconstrained system with $J=0.5$, as obtained from the
simulation, to be compared with the exact result obtained from the
diagonalization. }
\figure{Unconstrained polarization $|P|$, dielectric susceptibility $\chi$, and
specific heat $c_{v}$ for $J=1.,\; U=0$. Note the classical ferro-para
transition near $T_{c} \sim 1.$}
\figure{Finite size scaling determination of $T_{c}$, and
${{\gamma}\over{\nu}}$
from the phenomenological renormalization method \cite{barber}. On the vertical
axis is $$ f_{L_{1}L_{2}}^{\chi}(T) = {{\ln(\chi(L_{1},T)/\chi(L_{2},T))}\over
{\ln(L_{1}/L_{2})}} \; .$$ All the curves should intersect at the point
$(T_{c},{{\gamma}\over{\nu}})$. Our best estimate for $T_{c}$ is $1.0$, for
$J=1.,\; U=0$. }
\figure{Finite size scaling determination of $T_{c}$, and
${{\alpha}\over{\nu}}$, for $J=1.,\; U=0$.}
\figure{Polarization $|P|$, dielectric susceptibility $\chi$, and specific heat
$c_{v}$ for $J=0.75,\;U=0$. There still is evidence of a ferro-para transition
near $T_{c} \sim 0.5$. The singularity of $c_{v}$ is severely depressed by
quantum effects.}
\figure{Finite size scaling determination of $T_{c}$, and
${{\gamma}\over{\nu}}$, for $J=0.75,\; U=0$. Our best estimate is
$T_{c} \sim 0.5$.}
\figure{Dielectric susceptibility $\chi_{xx}$, and specific heat
$c_{v}$ for $J=0.5,\;U=0$. Long-range ferroelectricity is absent. On cooling,
the system evolves from classical paraelectric to QPE.}
\figure{Nearest-neighbour "short-range order parameter"
$\langle \cos(\phi_{ij} - \phi_{i^{'}j^{'}}) \rangle$ and polarization $|P|$
for $J=0.5,\;U=0$. Note a mild peak of the short-range order at
$T^{*} \sim 0.6$.}
\figure{Test of the PIMC method: dielectric susceptibility $\chi_{xx}$, and
internal energy $E$ for a $2 \times 2$ {\it constrained} system with
$J=0.5, \,U=\infty$,
as obtained from the simulation, to be compared with the exact result obtained
from the diagonalization. }
\figure{Constrained model dielectric susceptibility $\chi$ as a function of
coupling $J$, for $T=0.1$ and $T=0.2$, $U=\infty$. A ferro-para transition is
evident, near $J_{q}/t \simeq 0.15$.}
\figure{Polarization $|P|$, dielectric susceptibility $\chi$, and specific heat
$c_{v}$ for $J=0.3,\;U=\infty$. There is a clear ferro-para transition near
$T_{c} \sim 0.35$.}
\figure{Dielectric susceptibility $\chi_{xx}$, and specific heat
$c_{v}$ for $J=0.05,\;U=\infty$. The low-temperature state is a QPE.}
\figure{Nearest-neighbour "short-range order parameter"
$\langle \cos(\phi_{ij} - \phi_{i^{'}j^{'}}) \rangle$ and polarization $|P|$
for $J=0.05,\;U=\infty$. }
\figure{Phase diagram of both uncostrained and constrained model in the
$(J/t,T)$ plane. Note the shift of the phase boundary towards lower $J/t$ due
to the constraint.}

\begin{table}
\begin{tabular}{|p{6cm}|p{3cm}|p{3cm}|} \hline
  & $2 \times 2$ & $3 \times 3$ \\ \hline
$E_{g}^{exact}$ & -5.831 & -12.607 \\ \hline
$\langle \Psi_{g}^{\infty} | H |\Psi_{g}^{\infty} \rangle$ & -5.827 & -12.508
\\
\hline
$\langle \Phi_{g}^{exact}|\Psi_{g}^{\infty} \rangle$ & 0.9997 & 0.986 \\ \hline
$E_{exc}^{exact}$ & -4.903 & -11.906 \\ \hline
$\langle \Psi_{exc1}^{\infty} | H |\Psi_{exc1}^{\infty} \rangle$ & -4.866 &
-11.818  \\ \hline
$| \langle \Phi_{exc}^{exact}|\Psi_{exc1}^{\infty} \rangle |^{2} +
| \langle \Phi_{exc}^{exact}|\Psi_{exc2}^{\infty} \rangle |^{2}$ & - & 0.971
\\ \hline
$\langle \Phi_{exc1}^{exact}|\Psi_{exc1}^{\infty} \rangle$ & 0.995 & - \\
\hline
\end{tabular}
\caption[...]{\it Properties of the ansatz wavefunctions (\ref{ansatz}),
(\ref{excon}) as compared to exact wavefunctions for small systems,
$2 \times 2$ and $3 \times 3$.}
\label{t1}
\end{table}

\begin{table}
\begin{tabular}{|p{3cm}|p{3cm}|p{3cm}|} \hline
i & j & $\langle z^{*}_{i} z_{j}\rangle$ \\ \hline
(1,1) & (1,2) & 0.149 \\ \hline
(1,1) & (1,3) & $3. \times 10^{-2}$ \\ \hline
(1,1) & (1,4) & $(7.2 \pm 2.5) \times 10^{-3}$ \\ \hline
\end{tabular}
\caption[...]{\it Correlation function $\langle z^{*}_{i} z_{j}\rangle$ of the
ground state ansatz wavefunction (\ref{ansatz}) for a $10 \times 10$ system.}
\label{t2}
\end{table}


\begin{references}
\bibitem[36]{foot2} Since bonds cannot interchange in the present model, there
is no question as to their statistics, which is irrelevant. Physically,
however,
bonds are composite of (electron pair) + (lattice distortion), and can be
thought of as bosons.
\bibitem[41]{foot3} This will be possible in the future, provided
one can construct a suitable algorithm, built to take the constraint
automatically into account.
\end{references}
\end{document}